\documentclass[12]{amsart}
\usepackage{mathrsfs, amssymb, eucal, amsmath, amsthm, amscd, amsrefs}

\begin{document}
\pagestyle{headings}
\parindent0cm
\numberwithin{equation}{section}
\newtheorem{thm}{\bf Theorem}[section]
\newtheorem{lem}[thm]{\bf Lemma}
\newtheorem{prop}[thm]{\bf Proposition}
\newtheorem{cor}[thm]{\bf Corollary}
\newtheorem{defn}{\bf Definition}[section]
\theoremstyle{remark}
\newtheorem{rem}{\bf Remark}[section]
\newtheorem{exmp}{\bf Example}[section]
\title[The conservation of mass-moment parameters]{The conservation of mass-moment parameters}

\author[Dan Com\u{a}nescu]{Dan Com\u{a}nescu}
\address{Address of author: \\
Department of Mathemartics \\
West University of Timi\c{s}oara  \\
Bd. P\^{r}van nr. 4 \\
Timi\c{s}oara \\
Romania} \email{comanescu@math.uvt.ro}

\maketitle
\begin{abstract}
In this paper we study a concept of mass-moment parameter which is the generalization of \emph{the mass} and \emph{the moments of inertia} of a
continuous media. We shall present some interesting kinematical results in the hypothesis that a set of mass-moment parameters are conserved in
a motion of a continuous media.
\end{abstract}
\vspace{2mm}

\vspace{2mm}


\section{Introduction}\label{sec1}
\hspace{0,5cm} A mass-moment parameter of a continuous media is a kinematical parameter which is used to describe the distribution of the matter
in the continuous media. The most common mass-moment parameters are \emph{the mass} and \emph{the moments of inertia}. The mass-moment
parameters, with the exception of \emph{the mass}, are used especially in the theory of rigid bodies. In a motion of a rigid body the moments of
inertia are conserved with respect to a frame of reference rigidly connected to the body.

\hspace{0.5cm} The objective of our study are:
\begin{itemize}
    \item to present the concept of mass-moment parameter which is used in this paper;
    \item to introduce the conservation of a mass-moment parameter for a motion of a continuous media;
    \item to deduce the local laws in the cases in which one or two mass-moment parameters are conserved in a motion;
    \item to study the conservation of some moments of inertia.
\end{itemize}

\section{The mass-moment parameters}
\hspace{0.5cm} In this paper we consider the mass-moment parameters of the form (with respect to a frame of reference $\mathcal{R}$):
\begin{equation}\label{*}
    P(\mathcal{P},t)=\int_{P_t}\rho(\vec{x},t)p(\vec{x},t)dv
\end{equation}
where $t$ is a time-moment, $\mathcal{P}$ is a part of the continuous media, $P_t$ is the image of $\mathcal{P}$ (at time $t$), $\vec{x}$ is the
position vector of a particle (at time $t$), with respect to frame of reference $\mathcal{R}$, $\rho$ is \emph{the mass density} and $p$ is the
\emph{reduced density} of the mass moment parameter $P$.

\hspace{0.5cm} The mass-moment parameter $P$ is defined by a Lebesgue integral with respect to the volume measure. The functions $\rho$ and $p$
are supposed to be continuous.

\hspace{0.5cm} For the reduced density $p\equiv 1$ on obtain \emph{the mass} of the continuous media:
\begin{equation}\label{*}
    M(\mathcal{P},t)=\int_{P_t}\rho(\vec{x},t)dv
\end{equation}

The \emph{moment of inertia with respect to the point} $Q$ is defined by the integral:
\begin{equation}\label{*}
    I_Q(\mathcal{P},t)=\int_{P_t}\rho(\vec{x},t)d^2(\vec{x},\vec{x}_Q)dv
\end{equation}
where $d(\vec{x},\vec{x}_Q)$ is the distance between an arbitrary point of $P_t$ and $Q$.

The \emph{moment of inertia with respect to the plane} $\Pi$ is defined by the integral:
\begin{equation}\label{*}
    I_\Pi(\mathcal{P},t)=\int_{P_t}\rho(\vec{x},t)d^2(\vec{x},\Pi)dv
\end{equation}
where $d(\vec{x},\Pi)$ is the distance between an arbitrary point of $P_t$ and the plane $\Pi$.

The \emph{moment of inertia with respect to the $\Delta$-axis} is defined by the integral:
\begin{equation}\label{*}
    I_\Delta(\mathcal{P},t)=\int_{P_t}\rho(\vec{x},t)d^2(\vec{x},\Delta)dv
\end{equation}
where $d(\vec{x},\Delta)$ is the distance between an arbitrary point of $P_t$ and the line $\Delta$.

\section{The conservation of a set of mass-moment parameters}
\hspace{0.5cm} In this section we present some results concerning the conservation of one or two mass-moment parameters in a motion of a
continuous media.

First, we deduce the local laws for the conservation of a mass-moment parameter. Second, two conservation type results are deduced.
\bigskip

\begin{defn}
A mass-moment parameter $P$ is conserved in a motion of the continuous media if we have:
\begin{equation}\label{*}
    P(\mathcal{P},t_1)=P(\mathcal{P},t_2)
\end{equation}
for all parts $\mathcal{P}$ of the continuous media and for all two time-moments $t_1$ and $t_2$.
\end{defn}
\bigskip

We denote by $\vec{X}$ and $\vec{x}$ the position occupied by a particle at the initial moment and in the configuration at $t$. The law (3.1)
has the form:
\begin{equation}\label{*}
\int_{P_t}\rho(\vec{x},t)p(\vec{x},t)dv=\int_{P_0}\rho(\vec{X},0)p(\vec{X},0)dV
\end{equation}
\bigskip

\begin{rem}
The problem of conservation of a mass-moment parameter is connected with the mathematical theory of integral invariants.
\end{rem}
\bigskip

\begin{thm}
(\textbf{local law in material variables)} Let the $C^1$-map $\vec{x}=\chi(\vec{X},t)$ of the motion and $P$ a mass-moment parameter with $p$
the reduced density. The proposition are equivalents:

(i) $P$ is conserved in the motion;

(ii) for all time-moments $t$ and $\vec{X}\in D_0$ we have:
\begin{equation}\label{*}
\rho(\vec{x},t)p(\vec{x},t)J(\vec{X},t)=\rho(\vec{X},0)p(\vec{X},0)
\end{equation}
where $D_0$ is the image of the continuous media at the initial moment and $J$ the determinant of the deformation gradient; i.e.:
$$J(\vec{X},t)=det(\frac{\partial x_i}{\partial X_j}(\vec{X},t))_{i,j\in \{1,2,3\}}$$
\end{thm}
\bigskip

\begin{proof}
We need the following result:
\bigskip

\begin{lem}
(see [8] pp. 47) If $f:\mathcal{D}\subset \mathbb{R}^3\rightarrow \mathbb{R}$ is a continuous function on the domain $\mathcal{D}$ such that for
all domains $D\subset \mathcal{D}$ is satisfied $\int _D f dv=0$ then $f\equiv 0$ . \end{lem}
\bigskip

Using the relation (3.2) and the transformation of the integrals in the material variables we deduce that for all domains $P_0\subset D_0$ and
for all times $t$:
\begin{equation}\label{*}
\int_{P_0}\rho(\vec{x},t)p(\vec{x},t)J(\vec{X},t)-\rho(\vec{X},0)p(\vec{X},0)dV=0.
\end{equation}
 The theorem follows from (3.4) and Lemma 3.2.
\end{proof}
\bigskip

\begin{thm}
\textbf{ (local law in spatial variables)} Let the $C^2$-map $\vec{x}=\chi(\vec{X},t)$ of the motion and $P$ a mass-moment parameter with the
reduced density $p$ a $C^1$-function. The proposition are equivalents:

(i) $P$ is conserved in the motion;

(ii) for all time-moments $t$ and $\vec{x}\in D_t$ we have:
\begin{equation}\label{*}
\frac{d(\rho p)}{dt}(\vec{x},t)+(\rho p)(\vec{x},t)div_x\vec{v}(\vec{x},t)=0
\end{equation}
where $\vec{v}$ is the velocity and $div_x\vec{v}$ is the divergence of $\vec{v}$ with respect to the spatial variables.
\end{thm}
\bigskip

\begin{proof}
We need the Euler's lemma (see [8],pp. 36).

\begin{lem}
\textbf{(Euler)} For a $C^1$-motion $\vec{x}=\chi(\vec{X},t)$ of a continuous media is satisfied the relation:
$$\frac{dJ}{dt}(\vec{X},t)=J(\vec{X},t)div_x\vec{v}(\vec{x},t)."$$
\end{lem}

We have the sequence of equivalences:
$$(i)\,\,\,\Leftrightarrow\,\,\, \frac{d}{dt}\int_{P_t}\rho p \,dv=0\,\,\,\forall \mathcal{P}\,\,\,\Leftrightarrow\,\,\,\,\,\, \int_{P_0}\frac{dJ}{dt}\rho p+\frac{d\,\rho p}{dt}J \,dV=0\,\,\,\forall
\mathcal{P}$$ Using Euler's lemma and Lemma 3.2 we deduce:
$$(i)\,\,\,\Leftrightarrow\,\,\,\int_{P_t}\frac{d(\rho p)}{dt}+(\rho p)div_x(\vec{v},t)\,dv=0\,\,\,\forall \mathcal{P}\,\,\,\Leftrightarrow\,\,\,(ii)$$
\end{proof}
\bigskip

\hspace{0.5cm} We shall study now a case in which \emph{two} mass-moment parameters are conserved.
\bigskip

\begin{thm}
Let a $C^1$-motion  of a continuous media $\vec{x}=\chi (\vec{X},t)$. We consider two mass-media parameters $P_1$ and $P_2$ with reduced
densities $p_1$ and $p_2$ which are conserved in the motion. Then, for all particles and time-moments it is satisfied:
\begin{equation}\label{*}
    p_1(\vec{X},0)p_2(\vec{x},t)=p_1(\vec{x},t)p_2(\vec{X},0)
\end{equation}
\end{thm}
\bigskip

\begin{proof}
Following Theorem 3.1 one obtains:
$$\rho(\vec{x},t)p_1(\vec{x},t)J(\vec{X},t)=\rho(\vec{X},0)p_1(\vec{X},0)$$
$$\rho(\vec{x},t)p_2(\vec{x},t)J(\vec{X},t)=\rho(\vec{X},0)p_2(\vec{X},0)$$
Using the properties $\rho >0$ and $J\neq 0$ we deduce our result.
\end{proof}
\bigskip

\begin{thm}
If, moreover, $p_2\neq 0$ we deduce:
\begin{equation}\label{*}
    \frac{p_1}{p_2}(\vec{x},t)= \frac{p_1}{p_2}(\vec{X},0)
\end{equation}
\end{thm}
\bigskip

\begin{thm}
Let a $C^1$-motion  of a continuous media $\vec{x}=\chi (\vec{X},t)$ and $P$ a mass-moment parameter with $p$ the reduced density.

If the mass $M$ (see (2.2)) and $P$ are conserved in the motion then for all particles and time-moments we have:
\begin{equation}\label{*}
    p(\vec{x},t)=p(\vec{X},0)
\end{equation}
\end{thm}

\section{Applications}
\hspace{0.5cm} In this section are presented some applications of the results of section 3.

\subsection{Conservation of mass} If the mass $M$ (see (2.2)) is conserved in a $C^1$-motion $\vec{x}=\chi (\vec{X},t)$ of a continuous media
then, using Theorem 3.1, is obtained \emph{the material equation of continuity} (Euler 1762):
\begin{equation}\label{*}
    \rho (\vec{X},0)=\rho (\vec{x},t)J(\vec{X},t)
\end{equation}

If the motion is a $C^2$-function and $\rho$ is a $C^1$-function a consequence of conservation of mass is (see Theorem 3.3) \emph{the spatial
equation of continuity} (Euler 1757):
\begin{equation}\label{*}
    \frac{d\rho}{dt}(\vec{x},t)+\rho (\vec{x},t)div_x\vec{v}(\vec{x},t)=0
\end{equation}

\subsection{Conservation of mass and of an other mass-moment parameter} In this paragraph we suppose that a continuous media has a $C^1$-motion
such that the mass $M$ (see (2.2)) is conserved.

\hspace{0.5cm} Using Theorem 3.7 is easy to obtain the following results.
\bigskip

\begin{thm}
Let $O$ a fixed point. The following are equivalents:

\noindent (i) the moment of inertia with respect to the point $O$ is conserved in the motion;

\noindent (ii) the motion of an arbitrary particle is on a sphere with the center $O$.\bigskip
\end{thm}

\begin{thm}
Let $\Pi$ a fixed plane. The following are equivalents:

\noindent (i) the moment of inertia with respect to the plane $\Pi$ is conserved in the motion;

\noindent (ii) the motion of an arbitrary particle is in a plane parallel with $\Pi$.\bigskip
\end{thm}

\begin{thm}
Let $\Delta$ a fixed axis. The following are equivalents:

\noindent (i) the moment of inertia with respect to the $\Delta$-axis is conserved in the motion;

\noindent (ii) the motion of an arbitrary particle is on a circular cylinder with $\Delta$ as the axis of symmetry . \bigskip
\end{thm}

\subsection{A condition for the equilibrium of a continuous media} Let a $C^1$-motion of a continuous media with respect to a spatial frame of
reference $Ox_1x_2x_3$. We suppose that the mass $M$ is conserved in the motion. We denote by $I_O$ the moment of inertia with respect to the
point $O$, $I_{x_i}\,\, (i\in \{1,2,3\})$ the moment of inertia with respect to the $Ox_i$-axis, $I_{Ox_i x_j}\,\,\,(i,j\in \{1,2,3\})$ the
moment of inertia with respect to the plane $Ox_i x_j$. We have the relations (see [14] pp. 612-613):
\begin{equation}\label{*}
    I_O=\frac{1}{2}(I_{x_{1}}+I_{x_{2}}+I_{x_{3}})
\end{equation}

\begin{equation}\label{*}
    I_O=I_{Ox_1x_2}+I_{Ox_2x_3}+I_{Ox_3x_1}
\end{equation}

\begin{equation}\label{*}
    I_O=I_{x_i}+I_{Ox_jx_k},\,\,\,\{i,j,k\}=\{1,2,3\}
\end{equation}

\begin{equation}\label{*}
    I_{x_i}=I_{Ox_ix_j}+I_{Ox_ix_k},\,\,\,\{i,j,k\}=\{1,2,3\}
\end{equation}

\begin{equation}\label{*}
    I_{Ox_ix_j}=\frac{1}{2}(I_{x_i}+I_{x_j}-I_{x_{k}}),\,\,\,\{i,j,k\}=\{1,2,3\}
\end{equation}
\bigskip

\begin{thm}
If three of seven mass-moment parameters from the set\\
$I_O, I_{x_1}, I_{x_2}, I_{x_3}, I_{Ox_1x_2}, I_{Ox_2x_3}, I_{Ox_3x_1}$ are conserved in the motion then the continuous media is in an
equilibrium state.
\end{thm}
\bigskip

\begin{proof}
Using the relations (4.3)-(4.7) we deduce that all the seven mass-moment parameters are conserved in the motion of the continuous media.
Applying the Theorem 3.6 for the mass-moment parameters $I_{Ox_1x_2}, I_{Ox_2x_3}, I_{Ox_3x_1}$ and we obtain:
$$\left\{%
\begin{array}{ll}
    x_1^2(\vec{X},t)=X_1^2 \\
    x_2^2(\vec{X},t)=X_2^2 \\
    x_3^2(\vec{X},t)=X_3^2 \\
\end{array}%
\right.$$ The functions $t\rightarrow x_i(\vec{X},t)$ are continuous with the initial conditions $x_i(\vec{x},t)=X_i$. The result is
straightforward.
\end{proof}


\begin{bibdiv}
\begin{biblist}

\bib{Arnold74}{book}{title={Mathematical methods of classical mechanics}, author={Arnold, V.I.}, date={1974}, publisher={Nauka}, address={Moscow}}

\bib{Balint98}{book}{title={Lec\c{t}ii de mecanic\u{a} teoretic\u{a}. Mecanica solidului rigid}, author={Balint, St.}, date={1998}, publisher={Tip. Univ. de Vest}, address={Timi\c{s}oara}}

\bib{Balint96}{book}{title={Lec\c{t}ii de mecanic\u{a} teoretic\u{a}. Mecanica mediilor continue}, author={Balint, St.}, date={1996}, publisher={Tip. Univ. de Vest}, address={Timi\c{s}oara}}

\bib{Camenschi00}{book}{title={Introducere \^{i}n mecanica mediilor continue deformabile}, author={Camenschi, G.}, date={2000}, publisher={Ed. Univ. Bucure\c{s}ti}, address={Bucure\c{s}ti}}

\bib{Comanescu04}{book}{title={Modele \c{s}i metode \^{i}n mecanica punctului material}, author={Com\u{a}nescu, D.}, date={2004}, publisher={Mirton}, address={Timi\c{s}oara}}

\bib{Dragos03}{book}{title={Mathematical methods in Aerodynamics}, author={Drago\c{s}, L.}, date={2003}, publisher={Kluwer Academic Pub. and Ed. Academiei Rom\^{a}ne}, address={Bucure\c{s}ti}}

\bib{Dragos99}{book}{title={Mecanica fluidelor, vol. I. Teoria general\u{a}. Fluidul ideal incompresibil}, author={Drago\c{s}, L.}, date={1999}, publisher={Ed. Academiei Rom\^{a}ne}, address={Bucure\c{s}ti}}

\bib{Dragos83}{book}{title={Principiile mecanicii mediilor continue}, author={Drago\c{s}, L.}, date={1983}, publisher={Ed. Tehnic\u{a}}, address={Bucure\c{s}ti}}

\bib{Dragos76}{book}{title={Principiile mecanicii analitice}, author={Drago\c{s}, L.}, date={1976}, publisher={Ed. Tehnic\u{a}}, address={Bucure\c{s}ti}}

\bib{Iacob89}{book}{title={Matematic\u{a} aplicat\u{a} \c{s}i mecanic\u{a}}, author={Iacob, C.}, date={1989}, publisher={Ed. Academiei R.S.R.}, address={Bucure\c{s}ti}}

\bib{Iacob80}{book}{title={Mecanic\u{a} teoretic\u{a}}, author={Iacob, C.}, date={1980}, publisher={Ed. Didactic\u{a} \c{s}i Pedagogic\u{a}}, address={Bucure\c{s}ti}}

\bib{Soos83}{book}{title={Calcul tensorial cu aplica\c{t}ii \^{i}n mecanica solidelor}, author={So\'{o}s, E.}, author={Teodosiu, C. }, date={1983}, publisher={Ed. \c{S}tiin\c{t}ific\u{a} \c{s}i Enciclopedic\u{a}}, address={Bucure\c{s}ti}}

\bib{Truesdell74}{book}{title={Introduction a la Mecanique des Milieux Continus}, author={Truesdell, C.}, date={1974}, publisher={Mason}, address={Paris}}

\bib{Valcovici68}{book}{title={Mecanic\u{a} teoretic\u{a}, ed. a III-a}, author={V\^{a}lcovici, V.}, author={B\u{a}lan, St}, author={Voinea, R.}, date={1968}, publisher={Ed. Tehnic\u{a}}, address={Bucure\c{s}ti}}

\end{biblist}
\end{bibdiv}
\end{document}